# MAGENTICALLY-INDUCED LATTICE DISTORTIONS AND FERROELECTRICITY IN MAGNETOELECTRIC GdMnO$_3$

J. Agostinho Moreira[1], A. Almeida[1], J. Oliveira[1], M. R. Chaves[1], J. Kreisel[2], F. Carpinteiro[1] and P. B. Tavares[3]

[1] IFIMUP and IN-Institute of Nanoscience and Nanotechnology, Departamento de Física e Astronomia da Faculdade de Ciências da Universidade do Porto. Rua do Campo Alegre, 687, 4169-007 Porto, Portugal.

[2] Laboratoire Matériaux et Génie Physique, CNRS, Grenoble Institute of Technology, Minatec, 3, parvis Louis Néel, 38016 Grenoble, France

[3] Centro de Química. Universidade de Trás-os-Montes e Alto Douro. Apartado 1013, 5001-801. Vila Real. Portugal.

In this work we investigate the magnetic field dependence of $A_g$ octahedra rotation (tilt) and $B_{2g}$ symmetric stretching modes frequency at different temperatures. Our field-dependent Raman investigation at 10K is interpreted by an ionic displacive nature of the magnetically induced ferroelectric phase transition. The frequency change of the $A_g$ tilt is in agreement with the stabilization of the Mn-Gd spin arrangement, yielding the necessary conditions for the onset of ferroelectricity on the basis of the inverse Dzyaloshinskii-Moriya interaction. The role of the Jahn-Teller cooperative interaction is also evidenced by the change of the $B_{2g}$ mode frequency at the ferroelectric phase transition. This frequency change allows estimating the shift of the oxygen position at the ferroelectric phase transition and the corresponding spontaneous polarization of 480 µC.m$^{-2}$, which agrees with earlier reported values in single crystals. Our study also confirms the existence of a large magnetic hysteresis at the lowest temperatures, which is a manifestation of magnetrostiction.

Magnetically induced ferroelectricity has been reported in a variety of magnetoelectric systems and are considered of interest for both technical applications and fundamental understanding.[1-3] In a large number of systems, ferroelectricity has been understood on the ground of the inverse Dzyaloshinskii-Moriya (DM) model, where the modulated spin ordering induces lattice deformations, yielding electric polarization. In this scope, the interplay between the magnetic and polar degrees of freedom involves spin-lattice and thus spin-phonon coupling.[4-6] Spin-phonon coupling has been studied in several magnetoelectric materials by using both Raman and infrared spectroscopies[7-11] and have confirmed the significant role played by the DM mechanism in determining their phase diagrams. The DM mechanism is the proposed mechanism of magnetically induced polarization for some rare-earth manganites, like $R$MnO$_3$ with $R$= Gd, Tb, and Dy,[12] as well as, YMnO$_3$. Lattice dynamic studies have shown that the frequency of some phonons exhibit deviations regarding the pure anharmonic temperature behavior at the onset of the magnetic ordering, pointing to the coupling between phonon and magnetic excitations.[9,13] The emergence of ferroelectric ground states in rare-earth manganites is associated with the degree of orthorhombic distortion, which increases with decreasing the rare-earth ionic radius.[6] In TbMnO$_3$ and DyMnO$_3$, the ferroelectric ground state emerges even in the absence of an external magnetic field. In this case, the orthorhombic distortion is high enough to allow the antiferromagnetic exchange interactions to overcome the ferromagnetic ones, favouring the emergence of a commensurate spiral spin arrangement compatible with the stabilization of a ferroelectric phase. The situation is different for GdMnO$_3$, because of the large Gd$^{3+}$ ionic radius, but ferroelectricity can be induced by an external magnetic field.[12, 14-18] This fact is revealed by the (T,H) phase diagram obtained by Kimura *et al*[12] for GdMnO$_3$, under magnetic fields up to 9 T. In particularly, the phase boundary emerging at low temperatures for a magnetic field along the *b*-direction evidences a magnetic field-induced ferroelectric phase. This boundary is accompanied by a significant magnetic hysteresis signing a first order phase transition. The ferroelectric phase has been associated with the antiferromagnetic coupling between the Gd-4$f$ and Mn-3$d$ magnetic moments, and understood on the basis of symmetry arguments, where the canted antiferromagnetic Mn-spin arrangement presents a magnetic modulation vector of (0, 1/4 ,1).[12] Such a structure is non centrosymmetric, and is compatible with the emergence of a polarization along the *a*-direction, when a magnetic field is applied along the *b*-direction.[14]

a) Corresponding author. Email:

However, the origin of the magnetic field-induced ferroelectricity in GdMnO$_3$ is not yet fully understood. In particular, we are still lacking experimental data allowing favoring either ionic displacements or asymmetric electronic wave functions as a mechanism for stabilizing the ferroelectric phase.

In this scope, we present a detailed study of Γ-point phonons in GdMnO$_3$ under a magnetic field up to 6 T by using Raman scattering spectroscopy on high quality ceramics samples. Further details on ceramic processing and characterization, Raman spectra acquisition, and spectra analysis can be found in ref.[9,19]. In order to ensure reliable results, all of the Raman spectra were recorded at fixed holographic gratings positions. The thermal stability of the laboratory is better than 0.5 °K/day, while the thermal stability of the sample holder was better than 0.2 K. Moreover, Raman spectra were repeated several times at magnetic fields close to phase boundaries crossing, in order to obtain reliable data of the observed small changes.

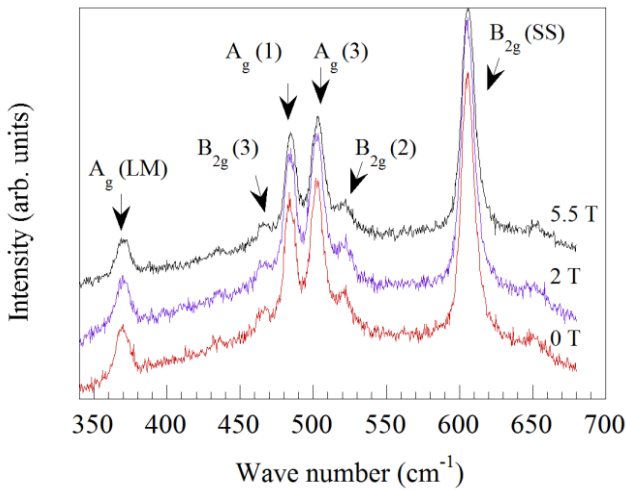

Figure 1: Raman spectra of GdMnO$_3$ recorded at 10 K, for several-fixed magnetic field strengths, in the 300-700 cm$^{-1}$ spectral range.

Figure 1 shows the Raman spectra of GdMnO$_3$ recorded at 10 K, for several-fixed magnetic field strengths, in the 300-700 cm$^{-1}$ spectral range. The mode assignment and the notation presented in Figure 1 are based on literature.[7,20,21] Although at the recorded temperature, a ferroelectric state is expected slightly above 1 T, as it is revealed from the (T,H) phase diagram obtained by Kimura *et al*,[12] both the profile and position of the bands show no important changes when the magnetic field is swept from 0 to 6 T. Namely, no new Raman-active bands emerge, which could have been taken as a direct sign for a magnetic field induced structural change. The absence of new Raman bands is, however, not surprising if we take into account both the reduced value of the oscillator strength of the polar modes at low temperatures and the improper nature of the ferroelectric phase.[2] In fact, the intensity of the activated Raman band is proportional to the oscillator strength, which is dependent on the TO-LO splitting. Nonetheless, a quantitative spectral analysis reveals for some of the Raman modes subtle but reliable anomalies in the magnetic field dependence of the mode parameters of both lattice and internal modes across the phase boundaries, which are the sign of a structural rearrangement thus subtle changes in atomic positions. Given the improper nature of the ferroelectric ground state, only very small atomic displacements are expected, in good agreement with previous structural studies carried out in several rare-earth manganites.[22,23]

Among all of the observed Raman-active modes, we have put particular attention to those carrying specific information of both lattice and molecular distortions, known to mirror the mechanisms underlying the magnetoelectric behavior in GdMnO$_3$. In the following, we will only focus on the $A_g$ lattice mode (LM) and the $B_{2g}$ symmetric stretching mode (SS) of the MnO$_6$ octahedra, marked in Figure 1. The former mode is assigned to an out-of-phase MnO$_6$ rotations around the [010] axis (P*bnm* setting) and has been associated with cooperative tilt vibrations of the octahedra.[7,20,21] This mode is thus a suitable local probe to sense lattice distortions associated with long range cooperative phenomena, like the emergence of lattice and/or magnetic modulated structures, which are most commonly found in certain magnetoelectric rare-earth manganites. Conversely, the Raman internal mode of $B_{2g}$ symmetry is a Jahn-Teller (JT) type mode, mainly associated with the (001) in-plane symmetric stretching of the O2-basal oxygens.[21] Alternating long and short MnO2 bonds in the *ab*-plane are a mark of the cooperative JT distortion. Since the Mn-O2 distances determine its frequency, this mode probes the relative octahedral distortions in the Mn-O2 plane, mainly associated with the Jahn-Teller mechanism.[20,21] Unravelling the role of the JT effect it plays in the behaviour of GdMnO$_3$ under high magnetic fields is of specific interest.

Figure 2 shows the magnetic dependence of the frequency of the $A_g$(LM) and $B_{2g}$(SS) modes in increasing and decreasing magnetic field runs at 50 K and 15 K. Since at these temperatures no phase line crossing is expected up to 6 T, as it follows from previous published results,[12] we should not observe any anomalies in the corresponding Raman spectra. In fact, for both temperatures the frequency of the $A_g$(LM) and $B_{2g}$(SS) modes reveals no anomalies as their frequency remains essentially independent on the magnetic field. This behaviour of both modes implies that the applied magnetic field does not yield significant changes of the structure and further distortions on the MnO$_6$ units. This feature is a direct consequence of the relative low strength of the spin-phonon coupling in these phases.[9,21]

We now focus on some fixed temperatures, where an anomalous magnetic behaviour of the modes parameters associated with the ferroelectric phase transition is expected. The (T,H) phase diagram obtained by Kimura et al[12] yields phase boundaries crossing at 10 K, and 5 K, while sweeping the magnetic field up to 6 T.

Figures 3a and 3b show the magnetic field dependence of the frequency of the $A_g$(LM) and $B_{2g}$(SS) modes in increasing and decreasing magnetic field runs at a fixed temperature of 10 K. For both modes, smeared anomalies are observed in the magnetic behaviour of their frequencies between 3T and 5 T, differing in their shape depending on the direction of the swept magnetic field. In order to understand these features it is worth to recall the magnetoelectric diagram from Kimura et al[12], where it can be observed that at 10 K just one type of boundary is crossed, corresponding to the ferroelectric phase transition with polarization along the a-direction induced by a magnetic field along the b-direction.

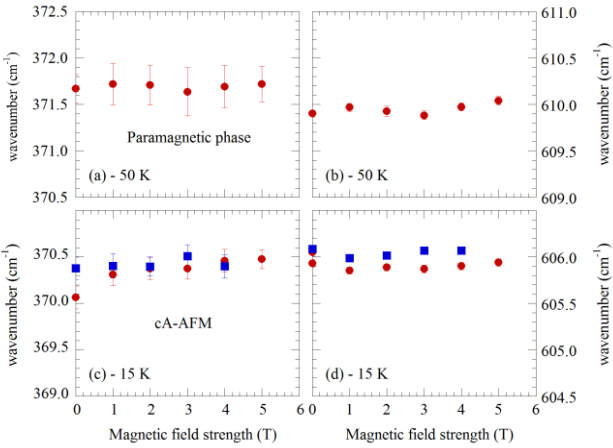

Figure 2: Magnetic field dependence of the frequency of the $A_g$(LS) and $B_{2g}$(SS) modes, at 50 K (a,b) and 15 K (c,d), in increasing (circles) and decreasing (squares) magnetic field runs.

A large magnetic hysteresis of about 2 T is observed. Despite a shift in the magnetic field, explained by the polycrystalline nature of the sample, the broad frequency steps and their common width of ~2 T evidence the transition from the canted A-type AFM to the ferroelectric phase, in good agreement with the phase diagram.[12] The $A_g$(LM) mode is associated with the tilt of the MnO$_6$ octahedra, its frequency increasing as the tilt angle increases.[7,20,21] The increasing frequency of the $A_g$ mode with magnetic field is thus a sign for an increase of the tilt angle which in turn leads to a weakening of the ferromagnetic exchange interaction. As a result, the balance of the competition between the ferromagnetic and antiferromagnetic exchange interactions alters, destabilizing the canted antiferromagnetic arrangement, with weak ferromagnetic component, and stabilizing a modulated spin structure, which is compatible with ferroelectricity through the inverse Dzyaloshinskii-Moriya interaction. The broad step observed in the magnetic field dependence of the frequency of the $A_g$ lattice mode reveals the stabilization of the Mn-3$d$/Gd-4$f$ spin arrangement with the magnetic modulation vector of (0,1/4,1).[12] We have estimated the relative change of the tilt angle associated with the modulated magnetic order by taking the change of the frequency of the $A_g$(LM) mode. For this, we assumed a linear relation between the tilt angle and the frequency of the $A_g$(LM) mode, which is expected from Landau theory and has been observed in rare-earth manganites.[7,21] Following this, the relative change of the tilt angle $\Delta\varphi$ can be expressed as:

$$\Delta\varphi/\varphi = \Delta\omega/\omega. \qquad (1)$$

From Figure 3a and increasing magnetic field strength, the increase of frequency of the $A_g$(LM) mode is 0.8 cm$^{-1}$, corresponding to a relative change of tilt angle of about 0.2% or, following Ref. 21, an absolute change of ≈ 0.05°. These values illustrate that the stabilization of the ferroelectric ground state involves very small atomic displacements, as will be also discussed further below. Another important issue is associated with the change in frequency of the $B_{2g}$(SS) mode observed above 3 T.

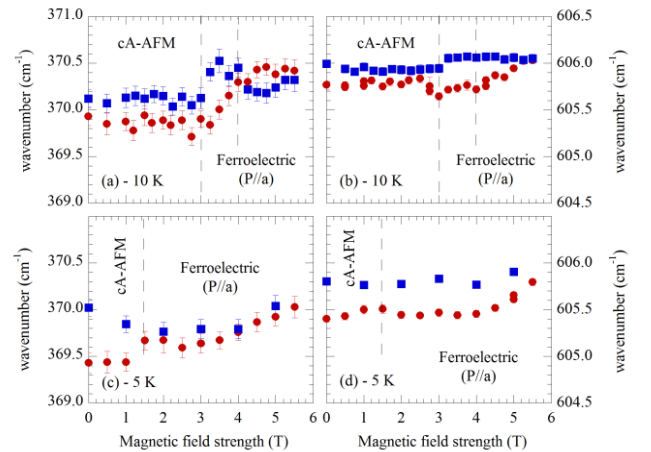

Figure 3: Magnetic field dependence of the frequency of the $A_g$(LS) and $B_{2g}$(SS) modes, at 10 K (a,b) and 5 K (c,d), in increasing (circles) and decreasing (squares) magnetic field runs.

The frequency of this mode, involving mainly stretching vibrations of O2 atoms in the ac plane, is determined by the Mn-O2 distances.[21] The change of frequency reflects the change in the Mn-O2 bond lengths.[21] Two conclusions can be drawn from the result displayed in Figure 3b. First, the emergence of the electric polarization is associated with the deformation of the MnO$_6$ octahedra, due to the displacement of the equatorial oxygen ions. As the Mn-O2 bonds lie mainly in the ac-plane, which is in agreement with a polarization along the a-direction as reported for single crystals. Moreover, the JT distortion seems to contribute to the stabilization of the ferroelectric phase. Second, we can estimate the order of magnitude of the polarization through the MnO2-length dependence of the $B_{2g}$(SS) mode frequency:[21]

$$\omega_{stretch} \propto d_{Mn-O2}^{3/2} \qquad (2)$$

In Equation (2) $d_{Mn-O2}$ is the mean value of the Mn-O2 bond length. The relative change in frequency due to changes on the Mn-O2 bond length is:

$$\frac{\Delta\omega_{stretch}}{\omega_{stretch}} = \frac{3}{2}\frac{\Delta d_{Mn-O2}}{d_{Mn-O2}} \qquad (3)$$

From Figure 3b we can get the value of the frequency shift due to the magnetically-induced phase transition. The frequency shift between 3 and 6 T is $\Delta\omega_{stretch} = 0.5$ cm$^{-1}$, while the value of the frequency for 3 T is $\omega_{stretch} = 605.6$ cm$^{-1}$. Taking $d_{Mn-O2} \approx 2$ Å from Ref. 22, we have calculated the corresponding displacement associated with the stabilization of the ferroelectric phase transition as ~0.00088 Å. Such a value could not be detected even by high resolution x-ray diffraction measurements.[22] Such a small displacement is also corroborated by the small increase in the tilt angle, as discussed above. In fact, and accordingly to earlier theoretical assumptions,[23] the increase of the tilt angle is intrinsically bound with the elongation of the Mn-O2 bonds in the *ac*-plane. For the estimation of the polarization we have then considered Z = 4, a unit cell volume of 230.5 Å$^3$, and the point charge model. Moreover, we assume the electric dipoles lie in the *ac*-plane. Under these assumptions, the estimated value for polarization is ≈480 μC.m$^{-2}$, close to the value of ≈400 μC.m$^{-2}$ reported in Ref. 12. The difference between the two values can emerge from the alignment of the electric dipoles, as it was assumed above.

Figures 3c and 3d show the magnetic field dependence of the frequency of the $A_g$(LM) and $B_{2g}$(SS) modes in increasing and decreasing magnetic field runs at a fixed temperature of 5 K. At this fixed temperature and from the phase diagram from Kimura *et al*,[12] a ferroelectric phase transition is expected at low magnetic fields. This transition is marked in the phonons frequencies at 1.5 T for increasing magnetic fields. Contrarily, for decreasing fields, their frequencies do not show any significant anomalies, which is in favour for a very large magnetic hysteresis, so large that the recovering of the canted A-type AFM phase cannot be achieved for finite temperatures, due to magnetostriction phenomena, in good agreement with the results in Ref. 16.

In summary, we have reported magnetic field induced changes in the phonon Raman spectra of GdMnO$_3$, which allow identifying the magnetically-induced ferroelectric phase transition in GdMnO$_3$. Our results are interpreted by considering polar atomic displacements, which could be associated with an asymmetric hybridization between the electronic wavefunctions of the rare-earth and oxygen ions.[24] A crucial role played by the oxygen ions in the emergence of the electric polarization is suggested.

Moreover, the importance role of both Jahn-Teller distortion and the modulation of the spin arrangement on the stabilization of the ferroelectric phase are distinctly evidenced. This behavior might well be a general feature in the orthorhombic rare-earth manganites exhibiting spiral spin order, which distinguish these compounds from the more distorted ones, exhibiting E-type spin order. Finally, the magnetic hysteresis observed at the lowest temperatures is a sign of magnetostriction and the first order nature of the transition into the ferroelectric phase.

**Acknowledgments.** This work was supported by Fundação para a Ciência e Tecnologia, through the Project PTDC/CTM/67575/2006.

1. W. Eerenstein, N. D. Mathur, and J. F. Scott, Nature (London) 442, 759 (2006).
2. Y. Tokura, J. Magn. Magn. Matter. 310, 1145 (2007).
3. A. B. Harris, A. Aharony, and O. Entin-Wohlman, J. Phys.: Condens. Matter 20, 434202 (2008).
4. T. Kimura and Y. Tokura, J. Phys.: Condens. Matter 20, 434204 (2008).
5. M. Kenzelmann, A. B. Harris, S. Jonas, C. Broholm, J. Schefer, S. B. Kim, C. L. Zhang, S.-W. Cheong, O. P. Vajk, and J. W. Lynn, Phys. Rev. Lett. 95, 087206 (2005).
6. M. Mochizuki and N. Furukawa, Phys. Rev. B 80, 134416 (2009).
7. J. Agostinho Moreira, A. Almeida, W. S. Ferreira, J. P. Araújo, A. M. Pereira, M. R. Chaves, J. Kreisel, S. M. F. Vilela, and P. B. Tavares, Phys. Rev. B 81, 054447 (2010).
8. Ch. Kant, J. Deisenhofer, T. Rudolf, F. Mayr, F. Schrettle, A. Loidl, V. Gnezdilov, D. Wulferding, P. Lemmens, and V. Tsurkan, Phys. Rev. B 80, 214417 (2009).
9. W. S. Ferreira, J. Agostinho Moreira, A. Almeida, M. R. Chaves, J. P. Araújo, J. B. Oliveira, J. M. Machado da Silva, M. A. Sá, T. M. Mendonça, P. Simeão Carvalho, J. Kreisel, J. L. Ribeiro, L. G. Vieira, P. B. Tavares, and S. Mendonça, Phys. Rev. B 79, 054303 (2009).
10. R. Valdés Aguilar, A. B. Sushkov, Y. J. Choi, S.-W. Cheong, and H. D. Drew, Phys. Rev. B 77, 092412 (2008).
11. T. Rudolf, Ch. Kant, F. Mayr, J. Hemberger, V. Tsurkan, and A. Loidl, Phys. Rev. B 76, 174307 (2007).
12. T. Kimura, G. Lawes, T. Goto, Y. Tokura, and A. P. Ramirez, Phys. Rev. B 71, 224425 (2005).
13. J. Laverdière, S. Jandl, A. A. Mukhin, V. Yu. Ivanov, V. G. Ivanov, and M. N. Iliev. Phys. Rev. B 73, 214301 (2006).
14. T. Arima, T. Goto, Y. Yamasaki, S. Miyasaka, K. Ishii, M. Tsubota, T. Inami, Y. Murakami, and Y. Tokura, Phys. Rev. B 72, 100102(R) (2005).
15. K. Noda, S. Nakamura, J. Nagayama, and H. Kuwahara, J. Appl. Phys.97, 10C103 (2005).
16. J. Baier, D. Meier, K. Berggold, J. Hemberger, A. Balbashov, J. A. Mydosh, and T. Lorenz, Phys. Rev. B 73, 100402(R) (2006).
17. R. Feyerherm, E. Dudzik, A. U. B. Wolter, S. Valencia, O. Prokhnenko, A. Maljuk, S. Landsgesell, N. Aliouane, L. Bouchenoire, S. Brown, and D. N. Argyriou, Phys. Rev. B 79, 134426 (2009).
18. T. Mori, N. Kamegashira, K. Aoki, T. Shishido, and T. Fukuda, Mater. Lett. 54, 238 (2002).
19. J. Agostinho Moreira, A. Almeida, W. S. Ferreira, M. R. Chaves, J. B. Oliveira, J. M. M. da Silva, M. A. Sá, S. M. F. Vilela, and P. B. Tavares, J. Electroceram. 25, 203 (2010).
20. L. Martín-Carrón, A. de Andrés, M. J. Martínez-Lope, M. T. Casais, and J. A. Alonso, Phys. Rev. B 66, 174303 (2002).
21. M. N. Iliev, M. V. Abrashev, J. Laverdière, S. Jandl, M. M. Gospodinov, Y.-Q. Wang, and Y.-Y. Sun, Phys. Rev. B 73, 064302 (2006).

22. J. Agostinho Moreira, A. Almeida, W. S. Ferreira, J. P. Araújo, A. M. Pereira, M. R. Chaves, M. M. R. Costa, V. A. Khomchenko, J. Kreisel, D. Chernyshov, S. M. F. Vilela, and P. B. Tavares, Phys. Rev. B 82, 094418 (2010).
23. M. Mochizuki, N. Furukawa, and N. Nagaosa, Phys. Rev. B 84, 144409 (2011).
24. C. D. Hu, Phys. Rev. B 77, 174418 (2008).